\documentclass{vgtc}                          

\graphicspath{{figures/}{pictures/}{images/}{./}} 

\usepackage{times}                     

\usepackage{tabu}                      
\usepackage{booktabs}                  
\usepackage{lipsum}                    
\usepackage{mwe}                       

\usepackage{mathptmx}                  

\usepackage{caption}
\usepackage{subcaption}

\onlineid{0}

\vgtccategory{Research}

\vgtcinsertpkg

\title{Nevis Digital Twin: Photogrammetry and Immersive Visualization of Historical Sites}

\author{Alex Apffel
\\ %
\scriptsize Department of Anthropology \\ \scriptsize San Jose State University, USA %
\and Huy Tran
\\ %
\scriptsize Department of Computer Science \\ 
\scriptsize San Jose State University, USA %
\and Vuthea Chheang\thanks{e-mail:vuthea.chheang@sjsu.edu}\\ 
\scriptsize Department of Computer Science \\ 
\scriptsize San Jose State University, USA}

\teaser{
  \centering
  \includegraphics[width=\linewidth]{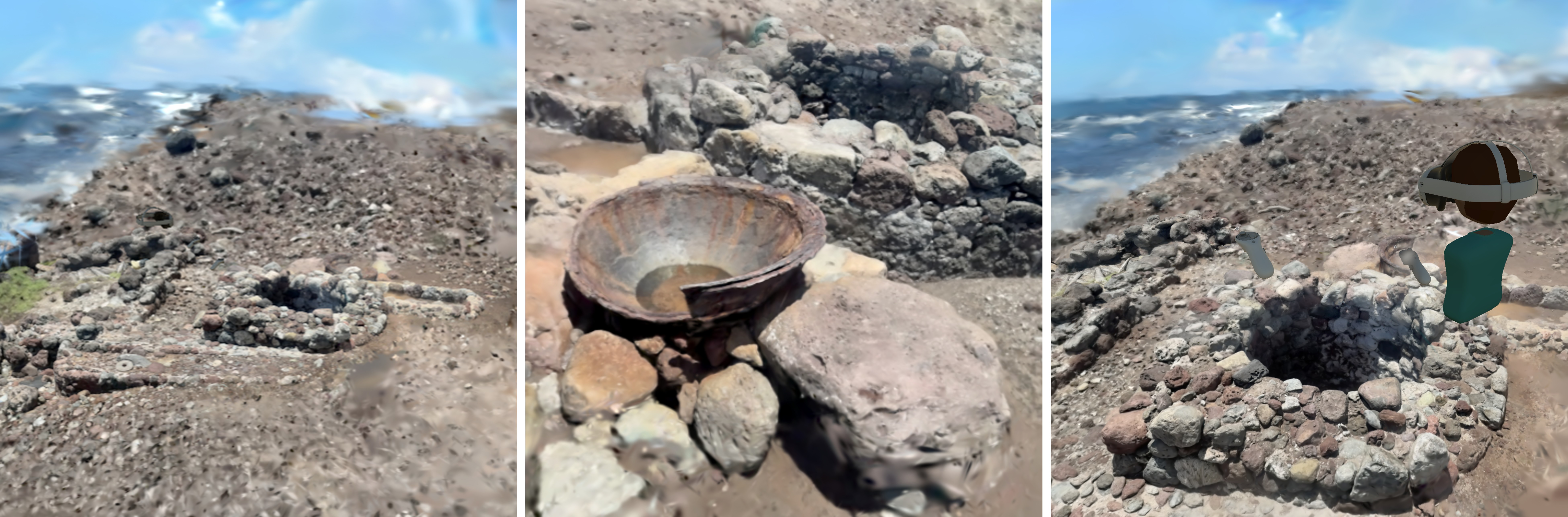}
  \caption{Virtual reconstruction of a well at White’s Bay on the island of Nevis in the Lesser Antilles: (a) overview of the 3D gaussian‑splatting environment generated from multimodal data acquisition; (b) view of the artifact and the well produced through the virtual reconstruction; and (c) VR visualization enabling immersive exploration and experience.}
  \label{fig:teaser}
}

\abstract{
    In this work, we present a multimodal data acquisition workflow for the digital preservation and virtual reconstruction of at-risk historical sites in the island of Nevis. Facing threats from coastal erosion, rising sea levels, and aggressive vegetation, the archaeological heritage of Nevis requires documentation strategies that bridge the gap between high-cost professional surveying and consumer accessibility.
    Experimental test compared acquisition variables, specifically camera height (1m vs. 3m) and operator trajectory against high-resolution control data.
    Moreover, we explore the virtual reconstruction between mesh reconstruction and 3D gaussian splatting to serve as different modalities for documentation. 
    The resulting data is fused into immersive virtual reality (VR) environments, offering a scalable, non-proprietary model for democratizing digital heritage in the Caribbean.
    
} 


\begin{document}


\firstsection{Introduction}

\maketitle


The island of Nevis in the Lesser Antilles, a pivotal location in the colonial history of the Caribbean, is home to a rapidly deteriorating landscape of historical significance. 
Prior to western contact, however, archaeological evidence suggests that Nevis was the site of contact between two separate cultures, the Taino and the Kalinago~\cite{nagele2020genomic}. 
The primary archaeological context for the present work is from this island of Nevis  where we are conducting studies to document cultural history of the Indigenous people who occupied the island prior to colonial contact. Between 3000 BCE and 1493 CE, the island was occupied by a succession of cultures of Guanahatabey, Taino and Kalinago people~\cite{dreyfustainos}. Both the Taino and Kalinago were seafaring people who also practiced slash-and-burn agriculture, growing crops such as cassava, yams, and plantains. They supplemented their diet with fishing and hunting, and their trading networks extended across the Caribbean. They lived in villages near the ocean and produced both pottery and stone tools and ritual objects. Following the discovery of St. Kitts and Nevis by Columbus on his second voyage, the arrival of European settlers led to conflicts with the Kalinago and the eventual displacement of the indigenous population. 

Over the last decade, a combination of natural erosion and climate change induced sea-level rise has led to the exposure of the prehistoric of settlements, artifacts and human remains from the Kalinago culture on the windward coast of Nevis. While this has uncovered previous unknown sites, the on-going process currently threatens the physical integrity of these areas and prompted efforts to rapidly document in as much detail as possible the status of the southeastern coast of the island with a combination of physical surface survey, and digital $360^o$ photographic landscape capture.

The use of photogrammetry for virtual reconstruction based on digital image capture provides archaeologists with an expanding toolbox across a range of scales, including 3D visualization of individual artifacts, interactive 3D documentation of excavation units, and GIS‑compatible 3D landscape reconstruction of entire study sites.
Ultimately, it would be desirable to combine all scales into single acquisitions, i.e., document study sites with sufficient resolution to identify individual artifacts. This data also offers new opportunities for archaeologist to explore, analyze and understand the environments they work in. Advances in machine learning and computer vision present the possibility to both identify and classify artifacts from these digital records for subsequent spatial and statistical analysis~\cite{dumonteil2025cognitive}. Virtual and Augmented Reality (VR/AR) technologies allow researchers to both share and re-interact with the site locations in unprecedented ways~\cite{sinha2025immersiveautochrome, gabel2025exploring, chheang2024enabling}. At the highest level, the goal of this collaborative research program is to explore ways in which new technologies present new opportunities. 
From the archaeological perspective, there are four primary goals for this project: 

\begin{itemize}
    \itemsep0em
    \item Establish a record for the current condition of the landscape in as much detail as possible to serve as a baseline for subsequent reevaluations. 
    \item Identify and prioritize specific areas for recovery and deeper investigation. 
    \item Produce quantitative spatial data relevant to Indigenous cultural studies. 
    \item Investigate and develop workflows for techniques for non-invasive image based landscape imaging and analysis. 
\end{itemize}

In addition to the field work in Nevis, a series of optimization studies were conducted in an artificial setting to study the effects of various image acquisition parameters, including camera type, camera direction, distance from ground and transect distance and speed. 

\section{Related Work}

Digital archiving the cultural heritages has increasingly emerged as a pivotal method to provide precise documentation and preservation as well as immersive visual representation~\cite{yu2025comparison}. 
Point cloud technology has been used in modern heritage digitization, which is a three-dimensional data representation captured the spatial morphology of objects based on their coordinates and color information of the surfaces. The approach for acquiring point cloud data includes LiDAR scanning and photogrammetry~\cite{owda2018methodology}.
Moreover, it serves as a primary tool in heritage digitization, enabling precise geometric capture, semantic modeling, and digital sharing. However, despite their accuracy, they struggle to deliver realistic surfaces, smooth real‑time rendering, and compelling visual narratives. 
These limitations have driven interest in newer approaches, such as image‑based neural rendering and Gaussian‑based methods, that can enhance or supplement point‑cloud‑based heritage visualization~\cite{jamil2025immersive}.

The image‑based neural rendering approaches often focus on generating realistic images by rendering a set of distinct geometric primitive and produce high-quality outputs based on those geometries~\cite{ kopanas2021point}. Nonetheless, in recent years the new implicit representation techniques has advanced and spurred researchers with neural implicit representation~\cite{fei20243d}. This method does not rely on any preset geometric structure to perform 3D reconstruction. 
3D gaussian splatting has rapidly replacing traditional meshing pipeline due to its ability to provide high-quality details, e.g., foliage and transparent glass, that are typically lost in the surface reconstruction~\cite{kerbl3Dgaussians, yu2025comparison}.
Thus, the shift from traditional photogrammetry to 3D gaussian splatting has created a need for new frameworks for data fusion and immersive visualization in archaeological heritage documentation.

\section{Material and Method}

\subsection{Data Acquisition}

We used the Insta360 X4 (Arashi Vision Inc., Shenzhen, China), a consumer-grade 360-degree camera, to overcome the field-of-view (FOV) limitations of standard DSLR photogrammetry. Traditional photogrammetry requires the operator to carefully aim the camera to ensure 60--80\% overlap between images. In contrast, a 360-degree camera captures the entire sphere of the environment simultaneously, rendering the \textit{aiming} process obsolete and allowing for a continuous walking trajectory.
Data acquisition was performed in three separate modes. 
\begin{itemize}
    \itemsep0em
    \item \textit{Timelapse Mode} (5.7K @ 2fps) using Insta360 X4 Camera with CupixVista/Vista Capture: This mode was tested to reduce data volume while maintaining high resolution. However, the lower frame rate can introduce gaps in the photogrammetric overlap if the operator moves too quickly.

    \item \textit{Video Capture Mode} (5.7K @ 30fps) with Insta360 X4 video capture: This became the preferred mode for the final workflow. Although it generates larger file sizes, the high frame rate ensures that redundant frames are available to eliminate motion blur and guarantee seamless ``Structure from Motion" (SfM) alignment.

    \item \textit{Frame Extraction} for individual building/structure and artifacts recorded using PolyCam (PolyCam Inc., San Francisco, California) on iPhone. 
\end{itemize}

For standard 3D landscape capture, the Insta360 X4 camera was operated at 5.7K resolution and 2fps acquisition rate under the control of the Vista Capture App (CupixVista, Seongnam-si, South Korea) on an iPhone. The camera held approximately 1m above head height on an invisible selfie stick. An area was covered at a normal walking pace in generally zig-zag pattern with approximately 2m transect spacing. 
A similar processes was repeated for a number of sites with the Insta360 X4 camera using 5.7K resolution and 30fps acquisition rate under the control of the Insta App (Insta360) on an iPhone. 
Individual buildings and artifacts were recorded using both iPhone and iPad cameras for photogrammetry. The acquisition control was with the PolyCam app on an iPhone and manual camera control with $>70\%$ image coverage. 
The workflow adapts construction industry ``Scan-to-BIM" tools for archaeological purposes.
\begin{itemize}
    \itemsep0em
    \item Frame Extraction: Raw 360-degree footage is processed in Insta360 Studio to stitch the dual-lens input into a single equirectangular video file.
    \item Sequence Generation: we extract individual JPEG frames from the video stream. The extraction rate is calculated based on walking speed to ensure a forward overlap of at least 70\%.
    \item Cloud Processing: The extracted frames are uploaded to CupixVista, a cloud-based 3D digital twin platform. CupixVista utilizes SLAM (Simultaneous Localization and Mapping) algorithms to align the 360-degree images, creating a 3D mesh and a navigable ``Street View" style tour.
    \item Artifact Modeling: For specific high-value objects identified during the survey, a secondary workflow using PolyCam on an iPad Pro (leveraging the LiDAR sensor and high-res camera) creates sub-millimeter accurate meshes that can be nested within the larger site model.
\end{itemize}

\subsection{Virtual Reconstruction}

\begin{figure}[t]
     \centering
     \begin{subfigure}[b]{0.46\columnwidth}
         \centering
         \includegraphics[width=\columnwidth]{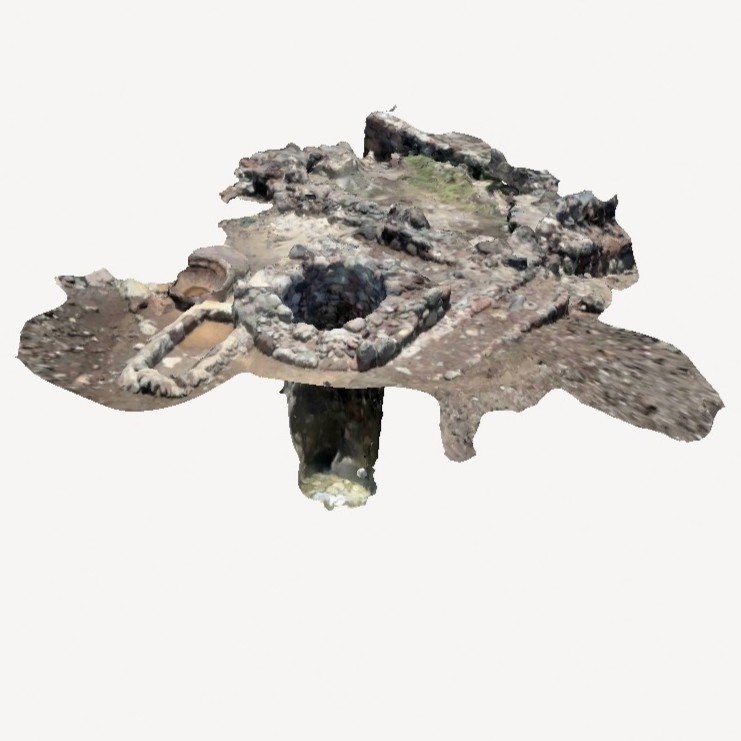}
         \caption{}
     \end{subfigure}
     \begin{subfigure}[b]{0.46\columnwidth}
         \centering
         \includegraphics[width=\columnwidth]{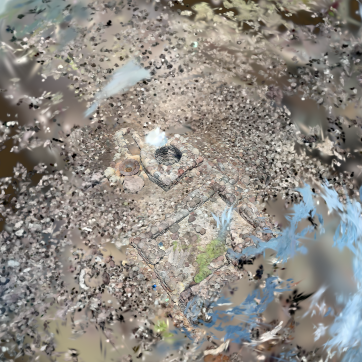}
         \caption{}
     \end{subfigure}
     \caption{Virtual reconstruction for a well at White's bay in the island of Nevis based on (a) mesh reconstruction and (b) 3D gaussian splatting.}
     \label{fig:virtualreconstruction}
\end{figure}

We aim to propose a digital twin namely ``Nevis Digital Twin" relies on fusing these photogrammetry datasets.
We explored two virtual reconstruction techniques based on meshes generated by PolyCam and 3D gaussian splatting (see~\autoref{fig:virtualreconstruction}). 

\paragraph{Mesh Reconstruction} 

The acquisition process follows an orbital capture methodology to ensure complete geometric coverage. In contrast to the linear path used for the landscape survey, the capture requires the operator to circumnavigate the target object. 
In this process, the iPad's LiDAR sensor is used, and the acquisition is performed in the three concentric loops: eye level, high angle, and low angle to minimize the occlusion. 
We use CupixVista to serves as the spatial skeleton. By importing the meshes generated by PolyCam, we can create a multi-scalar environment.

\begin{figure*}[t]
     \centering
     \begin{subfigure}[b]{0.23\textwidth}
         \centering
         \includegraphics[width=\textwidth]{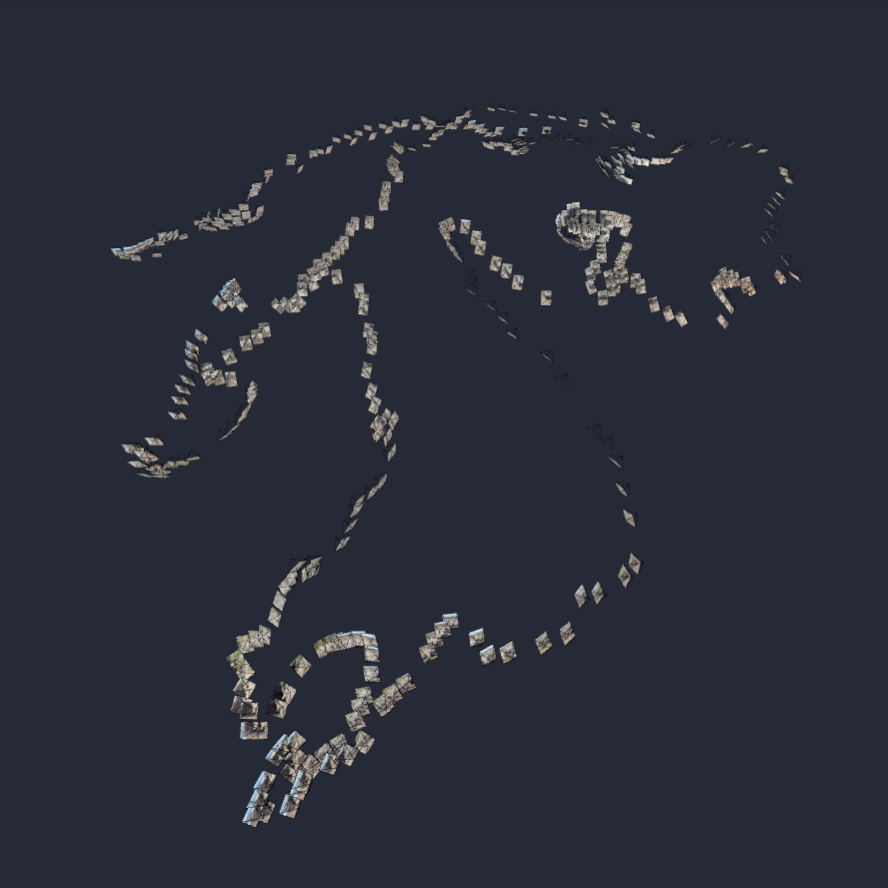}
         \caption{}
     \end{subfigure}
     \begin{subfigure}[b]{0.23\textwidth}
         \centering
         \includegraphics[width=\textwidth]{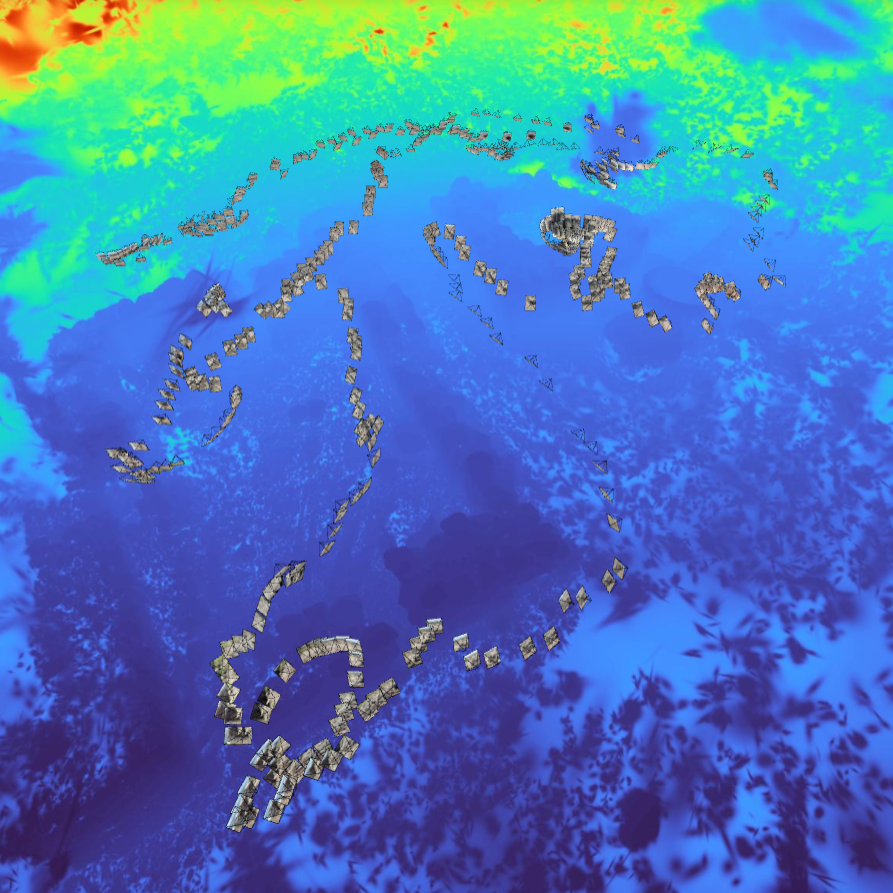}
         \caption{}
     \end{subfigure}
     \begin{subfigure}[b]{0.23\textwidth}
         \centering
         \includegraphics[width=\textwidth]{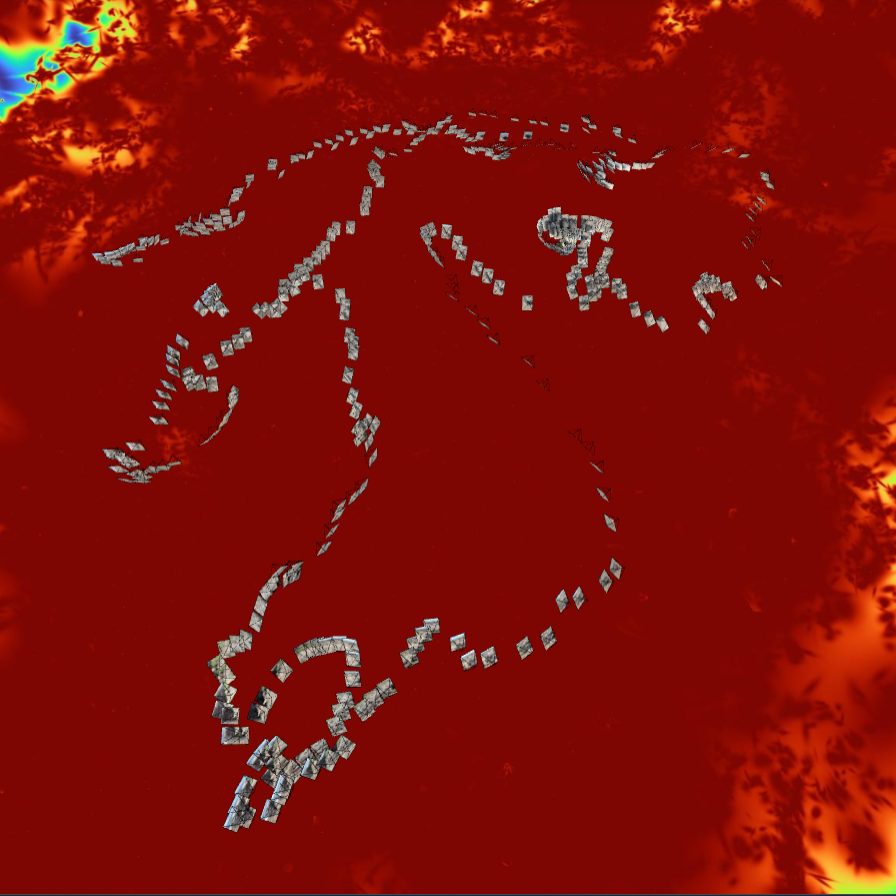}
         \caption{}
     \end{subfigure}
     \begin{subfigure}[b]{0.23\textwidth}
         \centering
         \includegraphics[width=\textwidth]{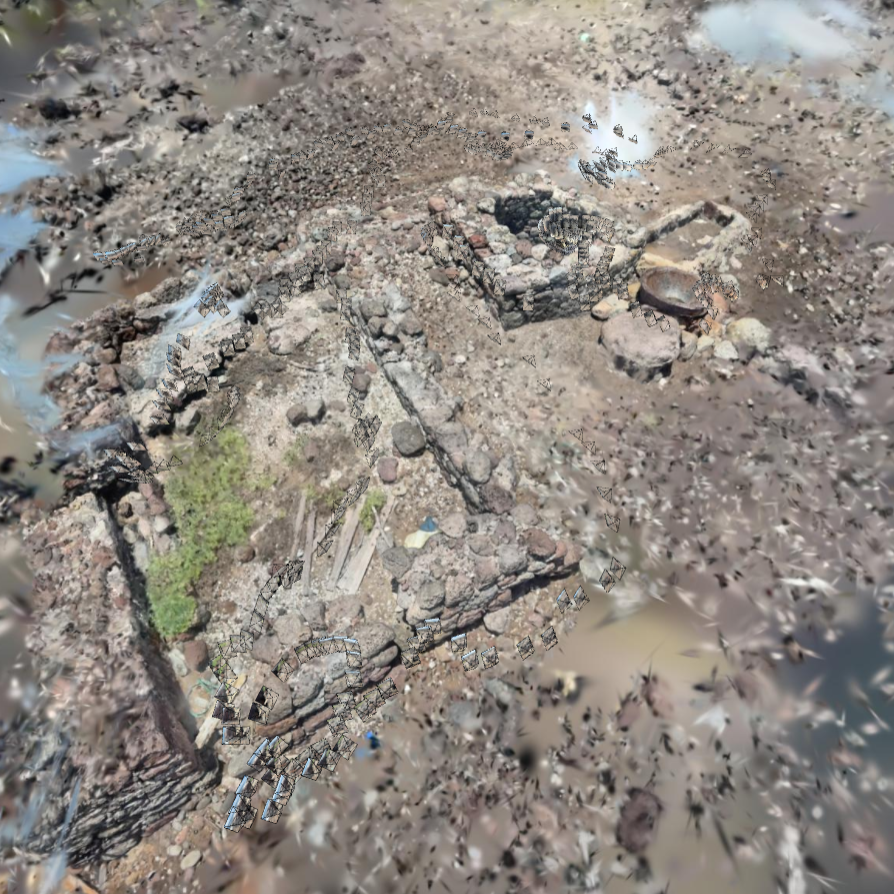}
         \caption{}
     \end{subfigure}
     \caption{Results of 3D gaussian splatting: (a) muti-view 2D images based on camera pose extracted from SfM algorithm, (b) depth view visualizing the distance from the camera for each pixel in the rendered image, (c) accumulation view summing color and density values along a ray to render pixels, and finally (d) the color view showing the final rendering.}
     \label{fig:gaussiansplat}
\end{figure*}

\paragraph{3D Gaussian Splatting}

While 3D meshes have been the standard for virtual reconstruction for decades, gaussian splatting can offer distinct advantages, especially for photorealistic reconstruction and complex scenes, which we believe provide strong potentials for the project.
We utilized the 3D gaussian splatting technique for real-time radiance field rendering proposed by Kerbl et al.~\cite{kerbl3Dgaussians}. 
The reconstruction pipeline was implemented using the Nerfstudio framework~\cite{nerfstudio}. We utilized the Splatfacto model, a streamlined implementation of 3D Gaussian Splatting designed for flexibility and performance. The model relies on \textit{gsplat} as its underlying Gaussian rasterization backend, which provides highly optimized CUDA kernels for accelerated differentiable rendering and rapid convergence. 

The virtual reconstruction using 3D gaussian splatting is a rasterization-based technique, which represents the scene as a collection of splats rather than geometric meshes. 
The workflow starts with a set of multi-view 2D images that can be extracted from the video capture using frame extraction technique (see~\autoref{fig:gaussiansplat}). 
We used a \textit{Structure-from-Motion} (SfM) algorithm, COLMAP~\cite{schoenberger2016sfm}, to estimate the camera pose from metadata for each image and generate a sparse 3D point cloud of the scene. This sparse point cloud is served as the initialization for the model.
Each point is this sparse point cloud is converted into a 3D gaussian primitive. Unlike simple point clouds, these 3D gaussian primitives are anisotropic and are defined by the learnable parameters, including position, covariance, opacity, and color.
The core of this virtual reconstruction is an iterative optimization loop. Thus, 3D gaussians are projected, in another word ``splatted'', onto the 2D image plane using a tile-based differentiable rasterizer. The resulting rendered image is then compared against the original ground-truth input images to calculate the loss (error). Finally, gradients are backpropagated to update the parameters (position, covariance, opacity, color) of every Gaussian.

\subsection{Virtual Reality (VR) Implementation}

The result of gaussian splatting was exported in a trained \textit{.PLY} file format as this format has become the standard interchange for gaussian splatting and due to its flexibility in storing data attributes. 
The real-time visualization of the 3D gaussian splatting was implemented in the Unity game engine (version 2022.3.62f2) using a custom rendering pipeline based on the open-source \textit{UnityGaussianSplatting} framework.
The trained PLY point clouds are imported and then converted into optimized custom Unity assets with selectable compression levels to minimize video memory usage.
To enable immersive VR visualization, we utilized Unity OpenXR and compute shaders provided by the open-source \textit{Unity VR Gaussian Splatting} plugin. 

The VR visualization was tested with Meta Quest 3 headset with $2064\times2208$ resolution per eye. 
We used a machine equipped an AMD Ryzen 9 5900HX with Radeon Graphics (16 CPUs)$@3.3GHz$ processor, RAM 32\,GB, and an Nvidia GeForce RTX 3070 (8\,GB VRAM) Laptop GPU for system development and testing.

\begin{figure}[t]
     \centering
     \begin{subfigure}[b]{\columnwidth}
         \centering
         \includegraphics[width=\textwidth]{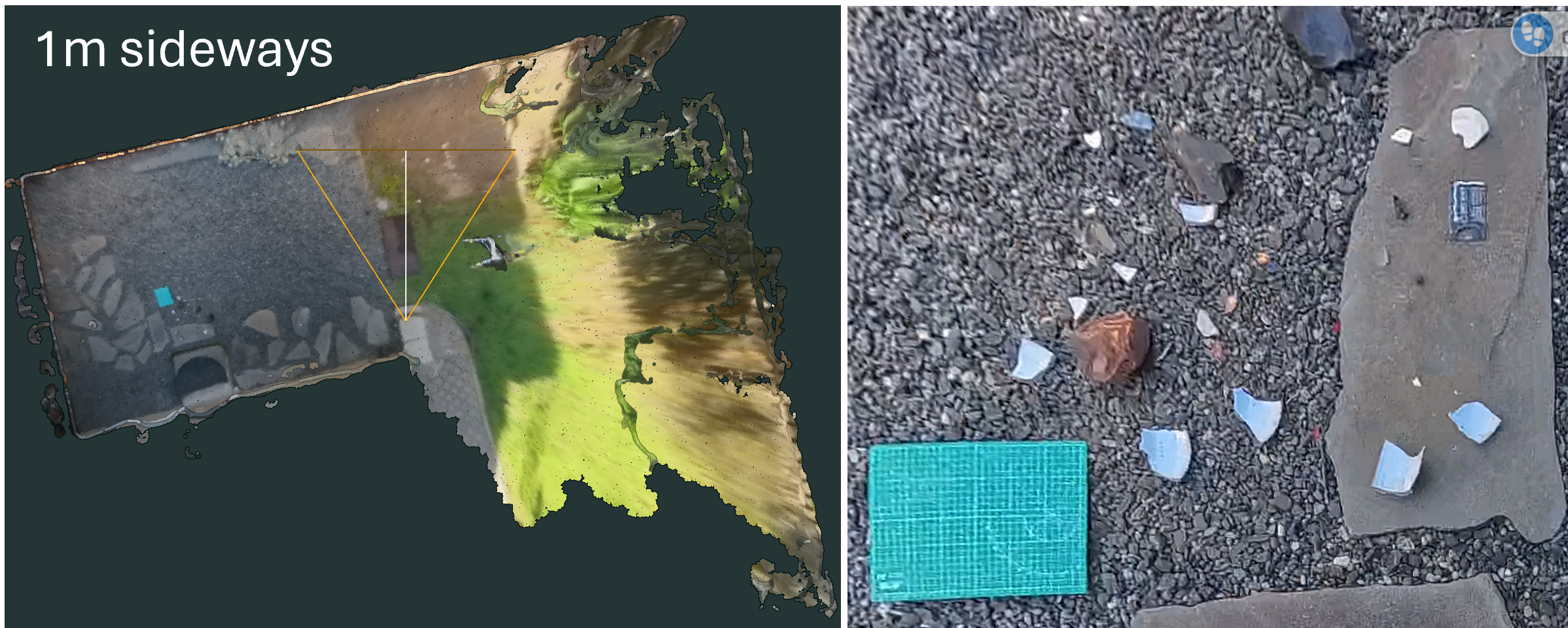}
         \caption{}
         \label{fig:cameraheight1}
     \end{subfigure}
     \begin{subfigure}[b]{\columnwidth}
         \centering
         \includegraphics[width=\textwidth]{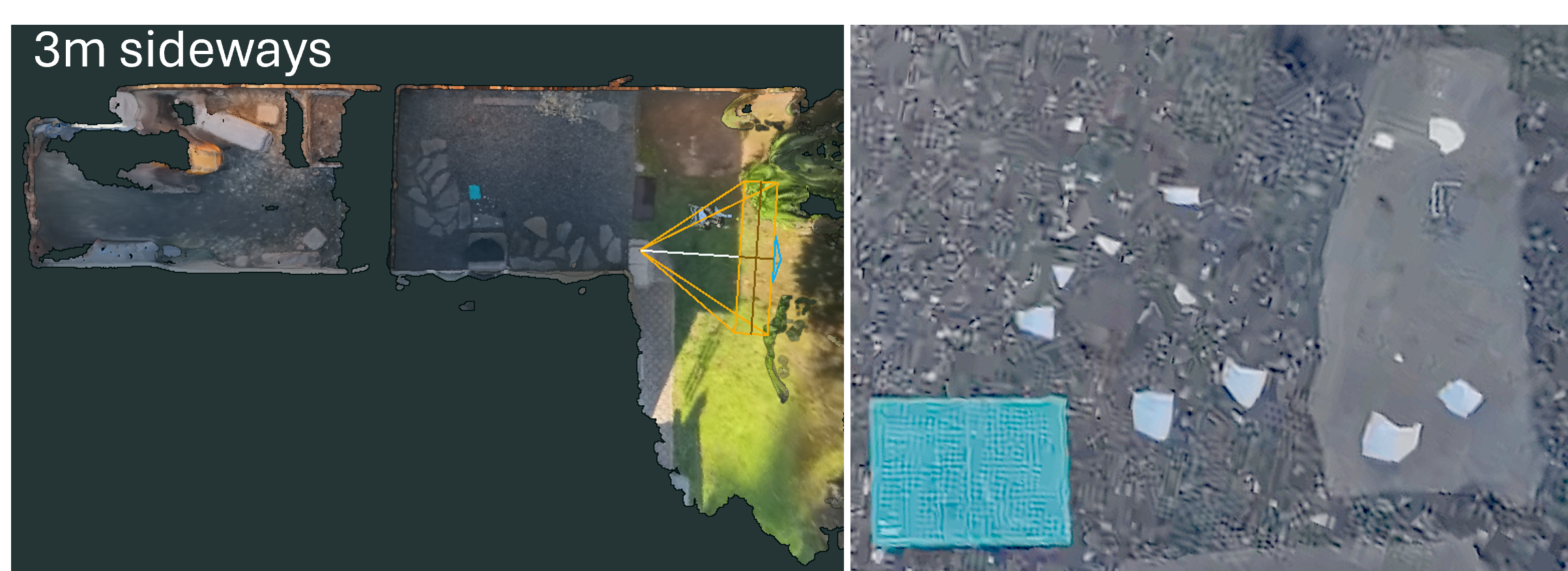}
         \caption{}
         \label{fig:cameraheight2}
     \end{subfigure}
     \caption{Effect of camera height using CupixVista: (a) 1‑meter sideways view and (b) 3‑meter sideways view. }
     \label{fig:cameraheight}
\end{figure}

\section{Results and Discussion}

\paragraph{Effects of various image acquisition parameters} 
A controlled test bed with approximately $5\times6$ meters was setup to study and validate the effects of image acquisition parameters with a typical archaeological surface scatter. 
A comparison between camera height 1\,m and 3\,m was conducted with the common assumption is that higher camera angles yield better mapping results for ground surveys (see~\autoref{fig:cameraheight}).
\begin{itemize}
    \item 1-meter acquisition (chest level): this configuration yielded superior results in CupixVista. The proximity to the ground maximized the effective resolution of the extracted frames. The study could confirm that 1-meter height is optimal for archaeological surveys where artifact identification, e.g., pottery and rock is required.
    \item 3-meter acquisition: The camera was mounted on an extended selfie stick with approximately 3 meters above the ground level. While this configuration provided a good overview of the site layout, the ground sampling distance for artifact identification was insufficient. The pixel density per centimeter was too low to resolve the texture of the artifact proxies.
\end{itemize}

Unlike drone photogrammetry, which is often blocked by the dense canopy of the Nevis rainforest, the handheld 360-workflow allows the operator to walk freely. The 360-degree field of view ensures that even in narrow paths, sufficient tie-points are captured on the ground and canopy ceiling to maintain model alignment.
The 1-meter acquisition height proves advantageous as it replicates a natural human eye-level perspective, increasing immersion compared to drone-based flyovers.

\begin{figure}[t]
     \centering
    \includegraphics[width=\columnwidth]{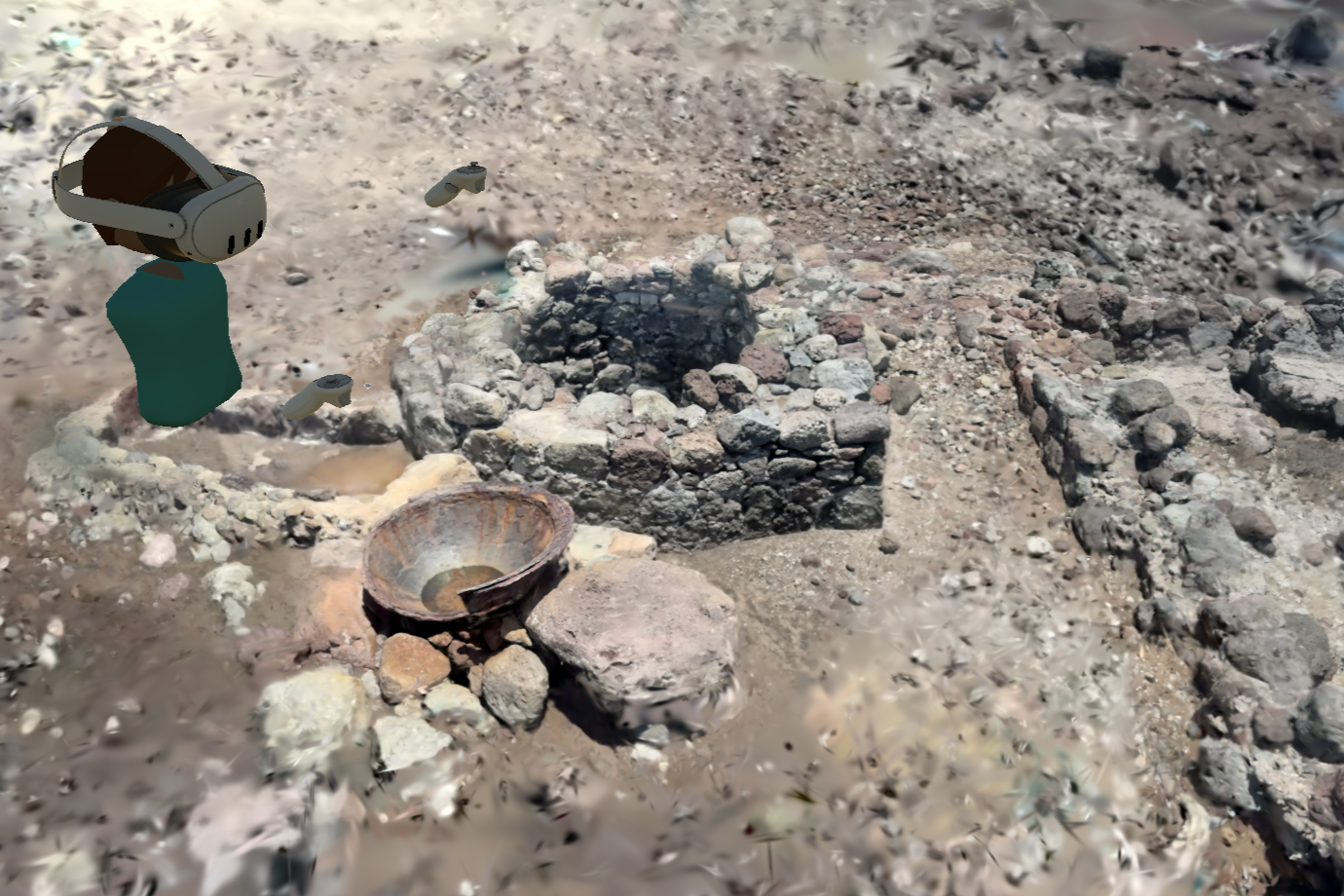}
     \caption{Real-time VR visualization of the virtual reconstruction with 3D gaussian splatting with a custom rendering pipeline.}
     \label{fig:xrvisualization}
\end{figure}

\paragraph{Virtual reconstruction and VR visualization}
Results of the virtual reconstruction based on 3D gaussian splatting on the use case of a well at White's bay in the island of Nevis are shown in \autoref{fig:gaussiansplat} and VR visualization in \autoref{fig:xrvisualization}. 
We used the same configuration and computer specification as mentioned in Sec. 3.3 to train the gaussian splatting model and use for real-time VR visualization. The trained model resulted in the number of splats around 953,425 with 225.5\,MB in size.

Conventional photogrammetry methods often struggle to capture fine structures such transparent materials and typically rely on heavy meshing, whereas 3D gaussian splatting can maintain high‑frequency detail by modeling scenes as volumetric collections of anisotropic 3D gaussians. It utilizes a tile-based rasterizer that sorts and projects splats significantly faster than NeRF ray-marching approaches. This speed is critical for VR, where maintaining high frame rates (typically 72Hz to 90Hz+ per eye) is essential to prevent motion sickness and ensure presence.
Despite its potential, deploying 3D gaussian splatting on standalone VR hardware, e.g., Meta Quest 3, Apple Vision Pro, introduces significant bottlenecks compared to PC-tethered setups. This includes memory bandwidth and VRAM constraints as 3D gaussian splatting scenes are memory-intensive, storage and streaming overhead since the file size of high-quality 3D gaussian splatting models often large, and aliasing and artifacts in VR because the users have 6 degree-of-freedom movement and can view splats from angles not covered in the training data. Thus, it could reveal artifacts or popping effects when the algorithm struggles with depth perception.

Future work will focus on the development of a robust pipeline for virtual reconstruction from 360-degree videos using 3D gaussian splatting. While current implementations largely rely on perspective images from pinhole cameras, 360-degree (omnidirectional) footage offers a unique advantage --- the ability to capture the entire radiance field of a scene from a single optical center, reducing the ``blind spots" that often cause holes in reconstruction.

Another challenge that we are focusing in the future is optimizing the trade offs between capture space,  capture speed and resolution. While 360-degree cameras provide excellent acquisition speed for large areas and comprehensive 360-degree coverage, distribution of the resolution capacity across the surface area of the spherical space dramatically reduced the optical image resolution compared with a conventional camera's field of view~\cite{chen2025splatter, byun2025crossgaussian}.  
Thus the 360-camera provides excellent reconstuction potential for landscape and architectural features, but may be challenged on smaller artifact identification. One continuing avenue of investigation is the optimization of hardware and acquisition parameters to produce sufficient ground level resolution while maintaining 360-degree spatial coverage in a single acquisition. 
Another approach is to utilize 360-degree acquisition for landscape and architecture reconstruction while conducting a separate overlapping groundlevel survey with a conventional camera for object identification and subsequently integrate and combine the data sets for a multimodal, multiscale data set. 


\section{Conclusion}

We present a \emph{Nevis Digital Twin}, workflow for digital documentation and virtual reconstruction of the at-risk historical sites in the island of Nevis.
Regarding data acquisition, the study concludes that for ground-based archaeological features, a 1-meter camera height using video-based extraction offers the optimal balance of speed and data utility. As sensor resolution improves in future 360-cameras, this workflow is poised to become the standard for documenting the endangered heritage.
In addition, we explore and present the virtual reconstruction and VR visualization with 3D gaussian splatting with a custom rendering pipeline on the use case of a well at White's bay.  
We have demonstrated that although computational and storage constraints currently limit standalone implementations, the trajectory of immersive technologies points toward a future in which photorealistic, volumetric digital twins become ubiquitous.


\acknowledgments{
We would like to acknowledge Prof. Marco Meniketti and the SJSU Nevis 2025 research team in supporting and enabling this work. 
This work was supported by the Department of Computer Science at San Jose State University through seed funding for conference registration.
}

\bibliographystyle{unsrt}

\bibliography{bibliography}
\end{document}